\documentstyle[12pt]{article}
\def\doit#1#2{\ifcase#1\or#2\fi}
\skewchar\fivmi='177
\skewchar\sixmi='177
\skewchar\sevmi='177
\skewchar\egtmi='177
\skewchar\ninmi='177
\skewchar\tenmi='177
\skewchar\elvmi='177
\skewchar\twlmi='177
\skewchar\frtnmi='177
\skewchar\svtnmi='177
\skewchar\twtymi='177
\def\@magscale#1{ scaled \magstep #1}

\def\framingfonts#1{
\doit{#1}{\font\twfvmi  = ammi10   \@magscale5 
\skewchar\twfvmi='177
\skewchar\fivsy='60
\skewchar\sixsy='60
\skewchar\sevsy='60
\skewchar\egtsy='60
\skewchar\ninsy='60
\skewchar\tensy='60
\skewchar\elvsy='60
\skewchar\twlsy='60
\skewchar\frtnsy='60
\skewchar\svtnsy='60
\skewchar\twtysy='60
\font\twfvsy  = amsy10   \@magscale5 
\skewchar\twfvsy='60
\font\go=font018			
\font\sc=font005			
\def\Go#1{{\hbox{\go #1}}}	
\def\Sc#1{{\hbox{\sc #1}}}	
\def\Sf#1{{\hbox{\sf #1}}}	
\font\oo=circlew10	      
\font\ooo=circle10			
\font\ro=manfnt				
\def\kcl{{\hbox{\ro 6}}}		
\def\kcr{{\hbox{\ro 7}}}		
\def\ktl{{\hbox{\ro \char'134}}}	
\def\ktr{{\hbox{\ro \char'135}}}	
\def\kbl{{\hbox{\ro \char'136}}}	
\def\kbr{{\hbox{\ro \char'137}}}	
}}

\catcode`@=11
\def\un#1{\relax\ifmmode\@@underline#1\else
	$\@@underline{\hbox{#1}}$\relax\fi}
\catcode`@=12

\let\du=\d			
\let\um=\H			

\def\a{\alpha}
\def\b{\beta}

\def\d{\delta}
\def\e{\epsilon}

\def\m{\mu}
\def\n{\nu}
\def\o{\omega}

\def\r{\rho}
\def\s{\sigma}

\def\L{\Lambda}

\def\plpl{{\raise-2pt\hbox{$\raise3pt\hbox{$_+$}\hskip-7.0pt\raise0.0pt
\hbox{$^+$}\hskip 0.01pt$}}}
\def\mimi{{\raise-2pt\hbox{$\raise3pt\hbox{$_-$}\hskip-7.0pt\raise0.0pt
\hbox{$^-$}\hskip 0.01pt$}}}

\def\bo{{\raise.15ex\hbox{\large$\Box$}}}		
\def\pr{\prod}						
\def\TH{{\raise.2ex\hbox{$\displaystyle \bigodot$}\mskip-4.7mu \llap H \;}}
\def\face{{\raise.2ex\hbox{$\displaystyle \bigodot$}\mskip-2.2mu \llap {$\ddot
	\smile$}}}					

\def\sp#1{{}^{#1}}				
\def\Tilde#1{{\widetilde{#1}}\hskip 0.03in}			
\def\Hat#1{\widehat{#1}}			
\def\Bar#1{\overline{#1}}			
\def\leftrightarrowfill{$\mathsurround=0pt \mathord\leftarrow \mkern-6mu
	\cleaders\hbox{$\mkern-2mu \mathord- \mkern-2mu$}\hfill
	\mkern-6mu \mathord\rightarrow$}
\def\dvec#1{\vbox{\ialign{##\crcr
	\leftrightarrowfill\crcr\noalign{\kern-1pt\nointerlineskip}
	$\hfil\displaystyle{#1}\hfil$\crcr}}}		
\def\dt#1{{\buildrel {\hbox{\LARGE .}} \over {#1}}}	

\def\frac#1#2{{\textstyle{#1\over\vphantom2\smash{\raise.20ex
	\hbox{$\scriptstyle{#2}$}}}}}			
\def\sfrac#1#2{{\vphantom1\smash{\lower.5ex\hbox{\small$#1$}}\over
	\vphantom1\smash{\raise.4ex\hbox{\small$#2$}}}}	
\def\bfrac#1#2{{\vphantom1\smash{\lower.5ex\hbox{$#1$}}\over
	\vphantom1\smash{\raise.3ex\hbox{$#2$}}}}	
\def\afrac#1#2{{\vphantom1\smash{\lower.5ex\hbox{$#1$}}\over#2}}    

\newskip\humongous \humongous=0pt plus 1000pt minus 1000pt
\def\caja{\mathsurround=0pt}
\def\eqalign#1{\,\vcenter{\openup2\jot \caja
	\ialign{\strut \hfil$\displaystyle{##}$&$
	\displaystyle{{}##}$\hfil\crcr#1\crcr}}\,}
\newif\ifdtup
\def\panorama{\global\dtuptrue \openup2\jot \caja
	\everycr{\noalign{\ifdtup \global\dtupfalse
	\vskip-\lineskiplimit \vskip\normallineskiplimit
	\else \penalty\interdisplaylinepenalty \fi}}}
\def\li#1{\panorama \tabskip=\humongous				
	\halign to\displaywidth{\hfil$\displaystyle{##}$
	\tabskip=0pt&$\displaystyle{{}##}$\hfil
	\tabskip=\humongous&\llap{$##$}\tabskip=0pt
	\crcr#1\crcr}}

\def\ref#1{$\sp{#1)}$}

\parskip=\medskipamount			
\thicklines			    

\def\oldheadpic{				
	\setlength{\unitlength}{.4mm}
	\thinlines
	\par
	\begin{picture}(349,16)
	\put(325,16){\line(1,0){4}}
	\put(330,16){\line(1,0){4}}
	\put(340,16){\line(1,0){4}}
	\put(335,0){\line(1,0){4}}
	\put(340,0){\line(1,0){4}}
	\put(345,0){\line(1,0){4}}
	\put(329,0){\line(0,1){16}}
	\put(330,0){\line(0,1){16}}
	\put(339,0){\line(0,1){16}}
	\put(340,0){\line(0,1){16}}
	\put(344,0){\line(0,1){16}}
	\put(345,0){\line(0,1){16}}
	\put(329,16){\oval(8,32)[bl]}
	\put(330,16){\oval(8,32)[br]}
	\put(339,0){\oval(8,32)[tl]}
	\put(345,0){\oval(8,32)[tr]}
	\end{picture}
	\par
	\thicklines
	\vskip.2in}
\def\oldtitle#1#2#3#4{\oldheadpic\begin{center}\vglue.5in{\large\bf #1}\\[.6in]
	{#2}\\[.1in] {\it Department of Physics and Astronomy}\\
	{\it University of Maryland, College Park, MD 20742}\\[.6in]
	Physics Publication \#{#3}\\ {#4}\\[1.5in] {\bf Abstract}\\[.1in]
	\end{center} \begin{quotation}}			
\def\oldTitle#1#2#3#4#5#6#7{\oldheadpic\begin{center} \vglue .4in
	{\large\bf #1}\\[.4in]
	{#2}\\[.1in] {\it Department of Physics and Astronomy}\\
	{\it University of Maryland, College Park, MD 20742}\\[.1in]
	{#3}\\[.1in] {\it {#4}}\\ {\it {#5}}\\[.4in]
	Physics Publication \#{#6}\\ {#7}\\[.5in] {\bf Abstract}\\[.1in]
	\end{center} \begin{quotation}}			
\def\border{						
	\setlength{\unitlength}{1mm}
	\newcount\xco
	\newcount\yco
	\xco=-24
	\yco=12
	\begin{picture}(140,0)
	\put(\xco,\yco){$\ktl$}
	\advance\yco by-1
	{\loop
	\put(\xco,\yco){$\kcl$}
	\advance\yco by-2
	\ifnum\yco>-240
	\repeat
	\put(\xco,\yco){$\kbl$}}
	\xco=158
	\yco=12
	\put(\xco,\yco){$\ktr$}
	\advance\yco by-1
	{\loop
	\put(\xco,\yco){$\kcr$}
	\advance\yco by-2
	\ifnum\yco>-240
	\repeat
	\put(\xco,\yco){$\kbr$}}
        \put(-20,11){\tiny University of Maryland Elementary Particle
Physics University of Maryland Elementary Particle Physics University of
Maryland Elementary Particle Physics}
	\put(-20,-241.5){\tiny University of Maryland Elementary
Particle Physics University of Maryland Elementary Particle Physics
University of Maryland Elementary Particle Physics}
	\end{picture}
	\par\vskip-8mm}
\def\bordero{						
	\setlength{\unitlength}{1mm}
	\newcount\xco
	\newcount\yco
	\xco=-24
	\yco=12
	\begin{picture}(140,0)
	\put(\xco,\yco){$\ktl$}
	\advance\yco by-1
	{\loop
	\put(\xco,\yco){$\kcl$}
	\advance\yco by-2
	\ifnum\yco>-240
	\repeat
	\put(\xco,\yco){$\kbl$}}
	\xco=158
	\yco=12
	\put(\xco,\yco){$\ktr$}
	\advance\yco by-1
	{\loop
	\put(\xco,\yco){$\kcr$}
	\advance\yco by-2
	\ifnum\yco>-240
	\repeat
	\put(\xco,\yco){$\kbr$}}
	\put(-20,12){\ooo
bacdefghidfghghdhededbihdgdfdfhhdheidhdhebaaahjhhdahbahgdedgehgfdiehhgdigicba}
	\put(-20,-241.5){\ooo
ababaighefdbfghgeahgdfgafagihdidihiidhiagfedhadbfdecdcdfagdcbhaddhbgfchbgfdacfediacbabab}
	\end{picture}
	\par\vskip-8mm}
\def\headpic{						
	\indent
	\setlength{\unitlength}{.4mm}
	\thinlines
	\par
	\begin{picture}(29,16)
	\put(165,16){\line(1,0){4}}
	\put(170,16){\line(1,0){4}}
	\put(180,16){\line(1,0){4}}
	\put(175,0){\line(1,0){4}}
	\put(180,0){\line(1,0){4}}
	\put(185,0){\line(1,0){4}}
	\put(169,0){\line(0,1){16}}
	\put(170,0){\line(0,1){16}}
	\put(179,0){\line(0,1){16}}
	\put(180,0){\line(0,1){16}}
	\put(184,0){\line(0,1){16}}
	\put(185,0){\line(0,1){16}}
	\put(169,16){\oval(8,32)[bl]}
	\put(170,16){\oval(8,32)[br]}
	\put(179,0){\oval(8,32)[tl]}
	\put(185,0){\oval(8,32)[tr]}
	\end{picture}
	\par\vskip-6.5mm
	\thicklines}
\def\title#1#2#3#4{\border\headpic {\hbox to\hsize{#4 \hfill UMDEPP #3}}\par
	\begin{center} \vglue .5in {\large\bf #1}\\[.6in]
	{#2}\\[.1in] {\it Department of Physics and Astronomy}\\
	{\it University of Maryland, College Park, MD 20742}\\[1.5in]
	{\bf Abstract}\\[.1in] \end{center} \begin{quotation}}	
\def\Title#1#2#3#4#5#6#7{\border\headpic
	{\hbox to\hsize{#7 \hfill UMDEPP #6}}\par
	\begin{center} \vglue .4in {\large\bf #1}\\[.4in]
	{#2}\\[.1in] {\it Department of Physics and Astronomy}\\
	{\it University of Maryland, College Park, MD 20742}\\[.1in]
	{#3}\\[.1in] {\it {#4}}\\ {\it {#5}}\\[.5in] {\bf Abstract}\\[.1in]
	\end{center} \begin{quotation}}			
\def\endtitle{\end{quotation}\newpage}			

\def\sect#1{\bigskip\medskip \goodbreak \noindent{\bf {#1}} \nobreak \medskip}
\def\refs{\sect{References} \footnotesize \frenchspacing \parskip=0pt}
\def\Item{\par\hang\textindent}

\def\doit#1#2{\ifcase#1\or#2\fi}
\def\[{\lfloor{\hskip 0.35pt}\!\!\!\lceil}
\def\]{\rfloor{\hskip 0.35pt}\!\!\!\rceil}

\def\Lag{{\cal L}}
\def\du#1#2{_{#1}{}^{#2}}
\def\ud#1#2{^{#1}{}_{#2}}
\def\dud#1#2#3{_{#1}{}^{#2}{}_{#3}}

\def\hati{{\hat{I}}}
\def\dt{$~D=10$~}

\def\pl#1#2#3{Phys.~Lett.~{\bf {#1}B} (19{#2}) #3}
\def\np#1#2#3{Nucl.~Phys.~{\bf B{#1}} (19{#2}) #3}

\def\pr#1#2#3{Phys.~Rev.~{\bf D{#1}} (19{#2}) #3}
\def\cqg#1#2#3{Class.~and Quant.~Gr.~{\bf {#1}} (19{#2}) #3}

\def\ap#1#2#3{Ann.~of Phys.~{\bf {#1}} (19{#2}) #3}

\def\ibid#1#2#3{{\it ibid.}~{\bf {#1}} (19{#2}) #3}

\def\ula{{\un a}}
\def\ulb{{\un b}}
\def\ulc{{\un c}}
\def\uld{{\un d}}

\def\fracmm#1#2{{{#1}\over{#2}}}
\def\gg{{\hbox{\sc g}}}
\def\half{{\fracm12}}

\def\frac#1#2{{\textstyle{#1\over\vphantom2\smash{\raise -.20ex
	\hbox{$\scriptstyle{#2}$}}}}}			

\def\fracm#1#2{\hbox{\large{${\frac{{#1}}{{#2}}}$}}}

\def\Dot#1{\buildrel{_{_{\hskip 0.01in}\bullet}}\over{#1}}
\def\dt#1{\Dot{#1}}
\def\uln{{\underline n}}
\def\Tilde#1{{\widetilde{#1}}\hskip 0.015in}
\def\Hat#1{\widehat{#1}}
\def\scst{\scriptstyle}
\def\itrema{$\ddot{\scriptstyle 1}$}
\def\Bo{\bo{\hskip 0.03in}}
\def\lrad#1{ \left( A {\buildrel\leftrightarrow\over D}_{#1} B\right) }
\def\derx{\partial_x} \def\dery{\partial_y} \def\dert{\partial_t}
\def\Vec#1{{\overrightarrow{#1}}}
\def\.{.$\,$}

\def\grg#1#2#3{Gen.~Rel.~Grav.~{\bf{#1}} (19{#2}) {#3} }
\def\pla#1#2#3{Phys.~Lett.~{\bf A{#1}} (19{#2}) {#3}}

\def\ula{{\underline a}}
\def\ulb{{\underline b}}
\def\ulc{{\underline c}}
\def\uld{{\underline d}}
\def\ule{{\underline e}}
\def\ulf{{\underline f}}
\def\ulg{{\underline g}}
\def\ulm{{\underline m}}
\def\uln#1{\underline{#1}}
\def\ulp{{\underline p}}
\def\ulq{{\underline q}}
\def\ulr{{\underline r}}

\def\hatm{\hat m}
\def\hatn{\hat n}
\def\hatr{\hat r}
\def\hats{\hat s}
\def\hatt{\hat t}

\def\ul{\underline}
\def\un{\underline}
\def\-{{\hskip 1.5pt}\hbox{-}}

\def\kd#1#2{\d\du{#1}{#2}}
\def\fracmm#1#2{{{#1}\over{#2}}}
\def\footnotew#1{\footnote{\hsize=6.5in {#1}}}

\def\low#1{{\raise -3pt\hbox{${\hskip 1.0pt}\!_{#1}$}}}

\def\ip{{=\!\!\! \mid}}
\def\unb{{\underline {\bar n}}}
\def\upb{{\underline {\bar p}}}
\def\um{{\underline m}}
\def\up{{\underline p}}
\def\Phib{{\Bar \Phi}}
\def\Phit{{\tilde \Phi}}
\def\Phibt{{\tilde {\Bar \Phi}}}
\def\Db{{\Bar D}_{+}}
\def\gg{{\hbox{\sc g}}}
\def\nt{$~N=2$~}

\def\Dot#1{\buildrel{_{_{\hskip 0.01in}\bullet}}\over{#1}}
\def\dt#1{\Dot{#1}}
\def\gg{{\hbox{\sc g}}}
\def\nt{$~N=2$~}
\def\gg{{\hbox{\sc g}}}
\def\nt{$~N=2$~}
\def\tr{{\rm tr}}
\def\Tr{{\rm Tr}}
\def\mpl#1#2#3{Mod.~Phys.~Lett.~{\bf A{#1}} (19{#2}) #3}
\def\hati{{\hat i}} \def\hatj{{\hat j}} \def\hatk{{\hat k}}
\def\hatl{{\hat l}}

\begin{document}

\font\tenmib=cmmib10
\font\sevenmib=cmmib10 at 7pt 
\font\fivemib=cmmib10 at 5pt  
\font\tenbsy=cmbsy10
\font\sevenbsy=cmbsy10 at 7pt 
\font\fivebsy=cmbsy10 at 5pt  
\def\BMfont{\textfont0\tenbf \scriptfont0\sevenbf
                              \scriptscriptfont0\fivebf
            \textfont1\tenmib \scriptfont1\sevenmib
                               \scriptscriptfont1\fivemib
            \textfont2\tenbsy \scriptfont2\sevenbsy
                               \scriptscriptfont2\fivebsy}
\def\rlx{\relax\leavevmode}
\def\BM#1{\rlx\ifmmode\mathchoice
                      {\hbox{$\BMfont#1$}}
                      {\hbox{$\BMfont#1$}}
                      {\hbox{$\scriptstyle\BMfont#1$}}
                      {\hbox{$\scriptscriptstyle\BMfont#1$}}
                 \else{$\BMfont#1$}\fi}

\font\tenmib=cmmib10
\font\sevenmib=cmmib10 at 7pt 
\font\fivemib=cmmib10 at 5pt  
\font\tenbsy=cmbsy10
\font\sevenbsy=cmbsy10 at 7pt 
\font\fivebsy=cmbsy10 at 5pt  
\def\BMfont{\textfont0\tenbf \scriptfont0\sevenbf
                              \scriptscriptfont0\fivebf
            \textfont1\tenmib \scriptfont1\sevenmib
                               \scriptscriptfont1\fivemib
            \textfont2\tenbsy \scriptfont2\sevenbsy
                               \scriptscriptfont2\fivebsy}
\def\BM#1{\rlx\ifmmode\mathchoice
                      {\hbox{$\BMfont#1$}}
                      {\hbox{$\BMfont#1$}}
                      {\hbox{$\scriptstyle\BMfont#1$}}
                      {\hbox{$\scriptscriptstyle\BMfont#1$}}
                 \else{$\BMfont#1$}\fi}

\def\inbar{\vrule height1.5ex width.4pt depth0pt}
\def\sinbar{\vrule height1ex width.35pt depth0pt}
\def\ssinbar{\vrule height.7ex width.3pt depth0pt}
\font\cmss=cmss10
\font\cmsss=cmss10 at 7pt
\def\ZZ{\rlx\leavevmode
             \ifmmode\mathchoice
                    {\hbox{\cmss Z\kern-.4em Z}}
                    {\hbox{\cmss Z\kern-.4em Z}}
                    {\lower.9pt\hbox{\cmsss Z\kern-.36em Z}}
                    {\lower1.2pt\hbox{\cmsss Z\kern-.36em Z}}
               \else{\cmss Z\kern-.4em Z}\fi}
\def\Ik{\rlx{\rm I\kern-.18em k}}  
\def\IC{\rlx\leavevmode
             \ifmmode\mathchoice
                    {\hbox{\kern.33em\inbar\kern-.3em{\rm C}}}
                    {\hbox{\kern.33em\inbar\kern-.3em{\rm C}}}
                    {\hbox{\kern.28em\sinbar\kern-.25em{\rm C}}}
                    {\hbox{\kern.25em\ssinbar\kern-.22em{\rm C}}}
             \else{\hbox{\kern.3em\inbar\kern-.3em{\rm C}}}\fi}
\def\IP{\rlx{\rm I\kern-.18em P}}
\def\IR{\rlx{\rm I\kern-.18em R}}
\def\IN{\rlx{\rm I\kern-.20em N}}
\def\Ione{\rlx{\rm 1\kern-2.7pt l}}

\def\sqrtuv{\sqrt{1-uv}}
\def\onemi{1-\fracmm1{r^4} }
\def\sinhth{\sinh\vartheta}
\def\coshth{\cosh\vartheta}
\def\tanhth{\tanh\vartheta}
\def\sqrtrf{{\sqrt{1-\fracmm1{r^4} }}}
\def\coshthsq{\cosh^2\vartheta}

\def\scst{\scriptstyle}
\def\itrema{$\ddot{\scriptstyle 1}$}
\def\Bo{\bo{\hskip 0.03in}}
\def\lrad#1{ \left( A {\buildrel\leftrightarrow\over D}_{#1} B\right) }
\def\derx{\partial_x} \def\dery{\partial_y} \def\dert{\partial_t}
\def\Vec#1{{\overrightarrow{#1}}}
\def\.{.$\,$}

\def\grg#1#2#3{Gen.~Rel.~Grav.~{\bf{#1}} (19{#2}) {#3} }

\def\pla#1#2#3{Phys.~Lett.~{\bf A{#1}} (19{#2}) {#3}}

\def\ula{{\underline a}} \def\ulb{{\underline b}} \def\ulc{{\underline c}}
\def\uld{{\underline d}} \def\ule{{\underline e}} \def\ulf{{\underline f}}
\def\ulg{{\underline g}} \def\ulm{{\underline m}}
\def\uln#1{\underline{#1}}
\def\ulp{{\underline p}} \def\ulq{{\underline q}} \def\ulr{{\underline r}}

\def\hatm{\hat m}\def\hatn{\hat n}\def\hatr{\hat r}\def\hats{\hat s}
\def\hatt{\hat t}

\def\plpl{{+\!\!\!\!\!{\hskip 0.009in}{\raise -1.0pt\hbox{$_+$}}
{\hskip 0.0008in}}}

\def\mimi{{-\!\!\!\!\!{\hskip 0.009in}{\raise -1.0pt\hbox{$_-$}}
{\hskip 0.0008in}}}

\def\items#1{\\ \item{[#1]}}
\def\ul{\underline}
\def\un{\underline}
\def\-{{\hskip 1.5pt}\hbox{-}}

\def\kd#1#2{\d\du{#1}{#2}}
\def\fracmm#1#2{{{#1}\over{#2}}}
\def\footnotew#1{\footnote{\hsize=6.5in {#1}}}

\def\low#1{{\raise -3pt\hbox{${\hskip 1.0pt}\!_{#1}$}}}

\def\ip{{=\!\!\! \mid}}
\def\unb{{\underline {\bar n}}}
\def\upb{{\underline {\bar p}}}
\def\um{{\underline m}}
\def\up{{\underline p}}
\def\Phib{{\Bar \Phi}}
\def\Phit{{\tilde \Phi}}
\def\Phibt{{\tilde {\Bar \Phi}}}
\def\Db{{\Bar D}_{+}}
\def\gg{{\hbox{\sc g}}}
\def\nt{$~N=2$~}

\def\framing#1{\doit{#1}
{\framingfonts{#1}
\border\headpic
\ifcase#1{\bf PRELIMINARY VERSION \hfill \today\\~~~}
\or {~~~}\fi
}}

\framing{0}

\vskip 0.03in

{\hbox to\hsize{December 1993\hfill UMDEPP 93--212}}\par

\begin{center}

{\large\bf Two--Dimensional ~Dilaton ~Gravity ~Black ~Hole ~Solution}\\
\vskip 0.01in
{\large\bf for}\\
\vskip 0.01in
{\large\bf $~N=2$~~ Superstring ~Theory}$\,$\footnote{This
work is supported in part by NSF grant \# PHY-91-19746.} \\[.1in]

\baselineskip 10pt

\vskip 0.25in

Hitoshi ~NISHINO\footnote{E-Mail: Nishino@UMDHEP.umd.edu} \\[.2in]
{\it Department of Physics} \\ [.015in]
{\it University of Maryland at College Park}\\ [.015in]
{\it College Park, MD 20742-4111, USA} \\[.1in]
and\\[.1in]
{\it Department of Physics and Astronomy} \\[.015in]
{\it Howard University} \\[.015in]
{\it Washington, D.C. 20059, USA} \\[.18in]

\vskip 0.5in

{\bf Abstract}\\[.1in]
\end{center}

\begin{quotation}

{}~~~We show that ~$N=8$~ {\it self-dual} supergravity theory, which is
the consistent background for $~N=2$~ closed superstring theory in
{}~$2+2\-$dimensions, can accommodate the recently discovered two-dimensional
dilaton gravity black hole solution, {\it via} appropriate dimensional
reductions and truncations.  Interestingly, the usual dilaton field in this
set of solutions emerges from the scalar field in the
$~{\bf 70}\-$dimensional representation of an intrinsic global $~SO(8)$~
group.  We also give a set of exact solutions, which can be interpreted
as the dilaton field on Eguchi-Hanson gravitational instanton background,
realized in an $~N=1$~ self-dual supergravity theory.
This suggests that the $~N=2$~ superstring has a close (even closer)
relationship with the two-dimensional black hole solution, which was
originally developed in the context of bosonic string and $~N=1$~ superstring.
Our result also provides supporting evidence for the conjecture that the
$~N=2$~ superstring theory is the ``master theory'' of supersymmetric
integrable systems in lower-dimensions.

\endtitle

\def\Dot#1{\buildrel{_{_{\hskip 0.01in}\bullet}}\over{#1}}
\def\dt#1{\Dot{#1}}
\def\gg{{\hbox{\sc g}}}
\def\nt{$~N=2$~}
\def\gg{{\hbox{\sc g}}}
\def\nt{$~N=2$~}
\def\tr{{\rm tr}}
\def\Tr{{\rm Tr}}
\def\mpl#1#2#3{Mod.~Phys.~Lett.~{\bf A{#1}} (19{#2}) #3}
\def\hati{{\hat i}} \def\hatj{{\hat j}} \def\hatk{{\hat k}}
\def\hatl{{\hat l}}

\oddsidemargin=0.03in
\evensidemargin=0.01in
\hsize=6.5in
\textwidth=6.5in

\centerline{\bf 1.~~Introduction}

There has been recently developing
interest in self-dual supersymmetric Yang-Mills and self-dual
supergravity (SDSG) theories in $~2+2\-$dimensions motivated by the
conjecture [1] that all the (bosonic)
integrable systems in lower dimensions can be generated by
self-dual Yang-Mills theory in four-dimensional space-time with
two time and two spacial coordinates.\footnotew{We denote this by $~D=(2,2)$,
where in
general $~D=(t,s)$~ denotes a space-time with $~t$~ time and $~s$~
spacial directions.  The expression $~D=4$~ is also used, when the
signature is not important.}\hfill Motivated by this development, we have
presented in our recent papers [2-6] convenient formulations for self-dual
supersymmetric Yang-Mills and  SDSG theories.  Remarkably, these self-dual
supersymmetric theories turned out to be also the consistent target space-time
backgrounds for $~N=2$~ open and closed superstring [7,8].  Among them, the
maximal SDSG theory has space-time $~N=8$~ supersymmetry [8] with
interesting features, such as the possibility of gauging a global
$~SO(8)$~ symmetry similar to the ordinary $~N=8$~ supergravity (SG) in
$~D=(1,3)$.

Rather independently of this subject in $~N=2$~ superstring, there has
been interesting development in ordinary bosonic string and $~N=1$~
superstring, about exact solutions for the  background dilaton and graviton
fields in two-dimensions [9].   It has been discovered that there
is a class of exact solutions similar to the usual four-dimensional
black hole solution that satisfies the background field equations
in those string theories [9].  These solutions are also identified with a
conformal sigma-model with a Wess-Zumino-Novikov-Witten (WZNW) term for
the coset $~SL(2,{\IR})/U(1)$~ [9].

In this paper, we show a new interesting link between these two
originally independent developments, namely the $~N=2$~ closed superstring
theory and the dilaton gravity black hole solutions.  We concentrate on the
$~D=(2,2),\,N=8$~ SDSG which is the consistent background for the closed
$~N=2$~
superstring, and apply to this system the technique of simple dimensional
reduction [10] into $~D=2$, in order to embed
the field equations of dilaton gravity black hole solutions.  Even though these
solutions have been developed as the  backgrounds for the ordinary bosonic or
$~N=1$~ superstring theories,  our result will establish the new viewpoint,
connecting it to the $~N=2$~ superstring.  Remarkably, we will see that the
usual dilaton emerges from the scalars in the $~{\bf 70}\-$dimensional
representation of the global $~SO(8)$~ symmetry in the  original $~N=8$~ SDSG
theory.   We also give another set of exact solutions, where the same
scalar field can be interpreted as the dilaton field in the
Eguchi-Hanson gravitational instanton background [11].

\vfill\eject

\centerline{\bf 2.~~Bosonic Field Equations for Ungauged $~N=8$~ SDSG
Background}

We start with reviewing the bosonic field part of the ungauged $~N=8$~
SDSG background field
lagrangian [8] for the $~N=2$~ closed superstring.
Its bosonic field content is the graviton (vierbein)
$~\Hat e\du{\hat\m} \hatm$, the
Lorentz connection $~\Hat\o\du{\hat\m}{\hatm\hatn}$, the propagating
multiplier field $~\Hat\L \du{\hat\m}{\hatm\hatn}$, the two
vector fields $\Hat A\low{\hat\m A B},~\Hat B\du{\hat\m}{A B} $~ each
in the $~{\bf 28}\-$representation of $~SO(8)$, and the scalar field
$~\Hat \phi\low{A B C D}$~ in the $~{\bf 70}\-$representation of the
$~SO(8)$.\footnotew{These fields
are respectively called $~e\low{\bf m}{}^{\a \a'},~\o\low{\bf m}{}^{\a\b}$,
$~\Tilde\o\low{\bf m}{}^{\a'\b'}, ~A\low{{\bf m}a b},~B\low{\bf m}{}^{a b}$,
and $~\phi_{a b c d}$~ in ref.~[8] in its own notation.  In our
notation, all the fields and indices with {\it hats} are for
$~D=4$, to be distinguished from $~D=2$~ quantities
later.  The indices $~{\scst \hat\m,~\hat\n,~\cdots~=~0,~\cdots,~3}$~
are for the $~D=4$~ curved indices, while $~{\scst
\hatm,~\hatn,~\cdots~=~(0),~\cdots,~(3)}$~ are for the local Lorentz
indices in $~D=4$.  Relevantly, we have the signatures
$~(\Hat\eta_{\hatm\hatn}) = \hbox{diag.}~(+,-,+,-)$~ for the
Minkowskian flat metric, and $~\Hat\e\, ^{0123} = + 1$.
This convention is slightly different
from that in our previous papers [2-6].  Additionally, the indices
$~{\scst A,~B,~\cdots~=~1,~\cdots,~8}~$ are for the ${\bf 8}$-representation of
$~SO(8)$ which correspond to $~{\scst a,~b,~\cdots}$~ in ref.~[8].  Our
anti-symmetrization symbols $~{\scst \[~\] }$~ are always normalized.}  The
purely bosonic  part of the lagrangian in the $~N=8$~ SDSG that interests us is
$$\eqalign{{\Hat\Lag}_{\rm B}^{N=8} & =
\,\half\Hat\e^{\,\hat\m\hat\n\hat\r\hat\s} \,
{\Hat e}_{\hat\m \hatm} \, \Hat\L
\du{\hat \n}{\hatm\hatn} \, \Hat T _{\hat\r\hat\s\hatn} + \fracm1{16}
{\Hat G}_{\hat\m\hat\n A B}\, {\Hat F}^{\hat\m\hat\n A B}  \cr
& - \fracm1{2304}\Hat e\, \e^{A B C D E F G H}
\left[ \, 2 \, \Hat g^{\hat\m\hat\n}  (\Hat\partial_{\hat\m} \Hat
\phi\low{A B C D}) (\Hat\partial_{\hat\n}  \Hat\phi\low{E F G H})
+ 3 \Hat \phi\low{A B C D}
\Hat F_{\hat\m\hat\n E F}^{(+)} \Hat F^{\hat\m\hat\n}_{(+) G H}
\,\right] ~~, \cr }
\eqno(2.1) $$
in an appropriate normalization of fields.
The $~\e^{A B C D E F G H}$~ is the totally antisymmetric constant invariant
tensor for $~SO(8)$.  If all the fermionic fields are ignored,
$$\Hat T_{\hat\m \hat\n}{}^{\hatm} \equiv
  \Hat\partial_{\hat\m} \Hat e \du{\hat\n}{\hatm}
- \Hat\partial_{\hat\n} \Hat e \du{\hat\m}{\hatm}
+ \Hat\omega\du{\hat\m}{\hatm\hatn} \Hat e_{\hat\n \hatn}
- \Hat\omega\du{\hat\n}{\hatm\hatn} \Hat e_{\hat\m \hatn}
\eqno(2.2) $$
is the usual torsion tensor with $~\Hat\omega\du{\hat\m}{\hatm\hatn}$~
regarded as an independent field [8].  This
$~\Hat\omega\du{\hat\m}{\hatm\hatn}$~
and the propagating multiplier field satisfy the self-duality and
anti-self-duality conditions, respectively:
$$\li{&\Hat\omega \du{\hat\m}{\hatm\hatn} = \half
\Hat\e\, \ud{\hatm\hatn}{\hatr\hats} \,\Hat\omega\du{\hat\m}{\hatr\hats} ~~,
&(2.3) \cr
& \Hat \L\du{\hat\m}{\hatm\hatn} = - \half \Hat\e \, \ud{\hatm\hatn}
{\hatr\hats} \, \Hat\L \du{\hat\m}{\hatr\hats} ~~.
&(2.4) \cr} $$
The $~\Hat F\low{\hat\m\hat\n A B}$~ and $~\Hat
G\du{\hat\m\hat\n}{A B}$~ are the field-strengths of the
vector fields $~\Hat A_{\hat\m A B}$~ and
$~\Hat B\du{\hat\m}{A B} $.  Since the $~SO(8)$~ symmetry is {\it global}, the
non-Abelian terms are absent in these field strengths.  The $~\Hat
G\du{\hat\m\hat\n}{A B}$~ satisfies the anti-self-duality condition: $$\Hat
G\du{\hatm\hatn}{A B} = - \half \Hat\e\, \du{\hatm\hatn}{\hatr\hats}  \Hat
G\du{\hatr\hats}{A B} ~~.
\eqno(2.5) $$
Eqs.~(2.3) - (2.5) are {\it not} field equations, but are built-in
conditions by definition, which become clearer in the spinorial notation in
ref.~[8].  The symbol $~{\scst(+)}$~ on $~\Hat F_{\hat\m
\hat\n A B}^{(+)}$~ denotes the self-dual part of
$~\Hat F_{\hat\m\hat\n A B}$.  Since $~\Hat\o\du{\hat\m}{\hatm\hatn}$~
is self-dual, the corresponding
Riemann tensor automatically satisfies the self-duality condition in this
formulation [8].  A comparison with the ordinary $~N=8$~ SG in $~D=(1,3)$
elucidates the peculiar property of this $~N=8$~ SDSG system due to the
propagating multiplier field $~\Hat\L\du{\hat\m}{\hatm\hatn}$, which maintains
the self-duality of the Lorentz connection,
while making the lagrangian formulation possible.  This special role
played by this multiplier field is similar to the ~$N=4$~ self-dual
supersymmetric Yang-Mills system [8], as described in our previous paper [5].

The field equations for the lagrangian, with all
fermionic fields suppressed, are
$$\li{&2\Hat\e\, ^{\hat\m\hat\n\hat\r\hat\s} {\Hat D}_{\hat\n} {\Hat\L}
_{\hat\r\hat\s\hatm} -\Hat\e\, ^{\hat\m\hat\n\hat\r\hat\s} \Hat
\L\du{\hat\n\hatm}\hatn \Hat T_{\hat\r\hat\s\hatn}  \cr
& ~~ + \fracm1{576} \Hat e\e^{A B C D E F G H} \left[\, \Hat e\du\hatm{\hat\m}
(\Hat D_{\hat\n} \Hat\phi\low{A B C D}) (\Hat D^{\hat\n} \Hat\phi\low{E
F G H}) - 2 (\Hat D_{\hatm}\Hat\phi\low{A B C D})(\Hat
D^{\hat\m}\Hat\phi\low{E F G H}) \right] \cr
& ~~ + \fracm1{384}\Hat e \e^{A B C D E F G H} \Hat\phi_{A B C D} \left[ \,
\Hat e\du\hatm{\hat\m} \Hat F_{\hat\r\hat\s E F}^{(+)}
\Hat F^{\hat\r\hat\s}_{(+) G H}
- 4 \Hat F_{\hatm\hat\s E F}^{(+)} \Hat F^{\hat\m\hat\s}_{(+) G H} \,\right] =
0
{}~~, {\hskip 0.6in}
&(2.6) \cr
&\Hat\e\, ^{\hat\m\hat\n\hat\r\hat\s} \left[\, \Hat e\du{\hat\n}{ \[ \hatm |}
\,\Hat T\du{\hat\r\hat\s}{| \hatn\]}
- \half \Hat\e\, \ud{\hatm\hatn}{\hatr\hats} \Hat e\du\n\hatr \Hat
T\du{\hat\r\hat\s}\hats \, \right] =0 ~~,
&(2.7) \cr
&\Hat\e\, ^{\hat\m\hat\n\hat\r\hat\s} \left[\,
\Hat \L\du{\hat\n\hat\r}{ \[ \hatm |}
\Hat e\du{\hat\s}{| \hatn \]} + \half \Hat\e\, \ud{\hatm\hatn}{\hatr\hats}
\Hat\L\du{\hat\n\hat\r} \hatr \hat e\du {\hat\s}\hats \,\right] =0 ~~,
&(2.8) \cr
& 4 \Hat D_{\hat\m}^2 \Hat\phi\low{A B C D} - 3 \Hat F_{\hat\m\hat\n
\[ A B|}^{(+)} \Hat F_{(+) | C D\]}^{ \hat\m\hat\n} = 0 ~~,
&(2.9) \cr
&\Hat D_{\hat\n} \Hat G^{\hat\m\hat\n A B} - \fracm1{24} \e^{A B C D E
F G H} \Hat D_{\hat\n}(\Hat\phi\low{C D E F} \Hat F^{\hat\m\hat\n}_{(+) G H} )
=
0~~,
&(2.10) \cr
&\Hat D_{\hat\n} \Hat F
\ud{\hat\m\hat\n}{A B} = 0 ~~,
&(2.11) \cr } $$
where $~\Hat D_{\hat\m}$~ is the general covariant derivative with
the Christoffel symbol $~\left\{ \dud{\hat\m}{\hat\r}{\hat\n} \right\}
$.  The familiar $~\pm 2$~ helicities of the graviton in the ordinary
{}~$D=(1,3)$~ are now shared by  $~\Hat\o\du{\hat\m}{\hatm\hatn}$~ and
$~\Hat\L\du{\hat\m}{\hatm\hatn}$, where especially the latter is a
propagating multiplier field, as seen in (2.6) and (2.7).
The last two lines in (2.6) are the source terms for the
$~\Hat\L\du{\hat\m}{\hatm\hatn}\-$field equation, whose important
significance will be seen later.

\bigskip\bigskip\bigskip

\centerline{\bf 3.~~Dilaton Gravity Black Hole Solution}

	We first give some ans\"atze for classical background solutions
to satisfy eqs.~(2.6) - (2.11).  First of all, we put simply
$$\Hat\L\du{\hat\m}{\hatm\hatn} = 0~~,
\eqno(3.1) $$
which satisfies (2.8) trivially, and makes the first two terms of (2.6)
vanish.  Note that it also obeys (2.4) as well.  In a way similar to the
simple dimensional
reduction scheme by Scherk-Schwarz [10], we specify the vierbein as
$$(\Hat e\du{\hat\m} \hatm) = \pmatrix{e^{-\varphi} ~e\du\m m & 0 \cr
0 & \d\du\a b \cr }~~,~~~~ (\Hat g\low{\hat\m\hat\n} ) = \pmatrix{e^{-2\varphi}
g\low{\m\n} & 0 \cr 0 & \eta\low{\a\b} \cr }~~,
\eqno(3.2) $$
where $~e\du\m m$~ and $~g_{\m\n}$~ are respectively the zweibein and
the metric in the resultant $~D=2$.  For our purpose of embedding the
black hole system [9], we have to further specify $~e\du\m m$~ and
$~g_{\m\n}$.  The simplest choice is
$$ (e\du\m m) = \fracmm 1{\sqrt 2}\pmatrix{e^{\varphi} & - e^{\varphi} \cr
e^{\varphi} & e^{\varphi} \cr } ~~, ~~~~
(g\low{\m\n}) = \pmatrix{0 & e^{2\varphi} \cr e^{2\varphi} & 0 \cr } ~~.
\eqno(3.3) $$
Here we are mainly using the notation in ref.~[10], namely
the curved indices $~{\scst \m,~\n,~\cdots~=~0,~1}$~ and the local
Lorentz indices $~{\scst m,~n,~\cdots~=~ (0),~(1)}$~ for $~D=(1,1)$~
with the Minkowskian metric $~(\eta_{m n})= \hbox{diag.}\,(+,-)$, into which
our dimensional reduction is performed, while the curved ones
$~{\scst \a,~\b,~\cdots~=~ 2,~3}$~ and the local Lorentz ones $~{\scst
a,~b,~\cdots~=~(2),~(3)}$~ for the {\it extra} two-dimensions
$~E=(1,1)$~ with the Minkowskian metric $~(\eta_{a b}) =
\hbox{diag.}\,(+,-)$.
All the fields with {\it hats} are in the original $~D=4$, while others {\it
without} hats are in the $~D=(1,1)$, to be distinguished from the former.
We have simply truncated the $~\Hat e\du\m\a\-$components just for
simplicity.  As usual in the simple dimensional reduction [10], we
require on all the fields their independence of the coordinates $~x^2$~ and
$~x^3$, namely $~\Hat\partial_\a=0$.
After this dimensional reduction, the original Lorentz symmetry $~SO(2,2)$~ is
eventually reduced to $~SO(1,1) \otimes SO(1,1)$.

Due to the opposite signs in the exponents in (3.2) and in (3.3), it is
clear that $~\Hat e\du{\hat\m}\hatm$~ is the {\it flat}
vierbein.\footnotew{Notice, however, that $~\Hat g_{\m\n}$~ does not
coincide with the $~SO(1,1)$-invariant metric $~\eta\low{m n}$, which can
be reached only after the coordinate transformation $~{x^0}' \equiv
(x^0+x^1)/{\sqrt2},~{x^1}' \equiv (x^0-x^1)/{\sqrt2}$.}  At first glance, this
sounds like eliminating all the physically important freedom of the system, but
as we will see shortly, this will result in highly non-trivial embedding of the
black hole solution into the $~D=4$~ SDSG system.

	   Because of the {\it flatness} of the $~D=4$~ metric,
we can simply put the ansatz for $~\Hat\o\du{\hat\m}{\hatr\hats}$~ as
$$\Hat\o\du{\hat\m}{\hat r\hat s} = 0 ~~,
\eqno(3.4) $$
satisfying also (2.3).  Accordingly we get
$$\Hat T\du{\hat\m\hat\n}{\hat m} = 0 ~~,
\eqno(3.5) $$
and we have now satisfied eq.~(2.7).  To solve the remaining
equations, we further set up our ans\"atze:\footnotew{The $\Hat
F_{\hat\m\hat\n A B}$~ in (3.7a) satisfies the self-duality condition.
However, we can also put non-zero anti-self-dual part, still satisfying all
the field equations.}
$$\li{&\Hat\phi_{1 2 3 4} = \phi(u,v) ~~,
&(3.6) \cr
&\Hat F_{m n \, 1 2} =\Hat F_{m n \, 3 4} = f \e_{m n}~~,
{}~~~~ \Hat F_{a b \, 1 2} = \Hat F_{a b \, 3 4} = - f \e_{a b} ~~,
&(3.7a) \cr
&\Hat G\du{\hat\m\hat\n}{A B} = 0 ~~,
&(3.7b) \cr } $$
and all other independent components of $~\Hat\phi\low{A B C D}$~ and
$~\Hat F_{\hat\m\hat\n A B}$~ are zero.  The $~f$~ is a constant to be
fixed later, and $~\e^{\m\n}$ and $~\e^{\a\b}$ are respectively the $~D=2$~
and $~E=2$~ Levi-Civita tensors such that $~\e^{01} = - \e^{10} = + 1,~
\e^{23} = - \e^{32}  = + 1$.  From now on, we use
$~u\equiv x^0,~ v\equiv x^1$~ for the $~D=2$~ coordinates.
Obviously (3.6) and (3.7) break the original
global $~SO(8)$~ symmetry down to a product of $~U(1)$.  Note also
that eq.~(3.7b) obeys (2.5), as desired.  These topological ans\"atze will
play an important role for a constant term in the ``dilaton'' equation, as we
will see later.

It is crucial to see how eq.~(2.6) is now satisfied, because
its last two lines left over after (3.1) are now vanishing due to
(3.6) and (3.7).  Eq.~(2.11) now holds because of (3.7).  The
satisfaction of (2.10) is also easily seen, because its last term is
vanishing under (3.6) and (3.7).  The only non-trivial equation left
over is (2.9) equivalent to
$$ \partial_u \partial_v \phi = \fracm 32 f^2 ~~,
\eqno(3.8) $$
which is nothing but the standard Poisson equation for the dilaton field in the
$~D=2$~ dilaton gravity black hole system [9].  By the appropriate
identification   $$\phi = e^{-2\varphi} ~~,
\eqno(3.9) $$
eq.~(3.8) is rewritten as
$$ 2 e^{-2\varphi} \left[ \, \partial_u \partial_v \varphi -2
(\partial_u\varphi) (\partial_v \varphi) \, \right] +\fracm32 f^2=0 ~~.
\eqno(3.10)$$
To accord with the familiar normalization [9], we put
$$ f = \fracm 2{\sqrt 3} ~~.
\eqno(3.11) $$

The only non-trivial consistency check now is the satisfaction of the $~D=2$~
gravitational equation for $e\du\m m$ in (3.3), because we are using
the same $~\varphi$~ both for (3.3) and (3.10).  In fact, by
computing the Ricci tensor in $~D=(1,1)$~ for (3.3), we get\footnotew{Note
that this Ricci tensor is computed purely in $~D=(1,1)$, which should {\it
not} be confused with $~\Hat R\du\m\n$~ in $~D=(2,2)$, {\it etc.}}
$$ R_{\m\n} = g_{\m\n} D_\r ^2 \varphi = 2 D_\m D_\n
\varphi  ~~,
\eqno(3.12) $$
which coincides with the dilaton gravity black hole
gravitational  equation [9].  The first equality can be understood as the
natural result for the Weyl rescaling for a metric in $~D=2$~ as in (3.2).  The
second equality is due to the relations $~D_0^2 \varphi = D_1^2
\varphi=0$~ for (3.3).  Eq.~(3.10) with
(3.11) can be put into a general covariant form, by the use of the relation
$~D_\r^2 \varphi = 2 e^{-2\varphi} \partial_u \partial_v \varphi$~ under (3.3):
$$D_\m^2 \varphi - 2(D_\m\varphi)^2 + 2 = 0 ~~.
\eqno(3.13)$$
After all, our independent covariant field equations are
(3.12) and (3.13), coinciding with the familiar dilaton gravity black
hole system, which has the solutions [9]
$$\li{& (g\low{\m\n}) = \fracmm 12
\fracmm1{u v-1} \pmatrix{0 & 1 \cr  1 & 0 \cr} ~~, ~~~~
(e\du\m m) = \fracmm 12 \fracmm 1{\sqrt{u v-1}}
\pmatrix{1 & - 1
\cr 1 & 1\cr } ~~,
&(3.14) \cr
&\varphi = -\half \ln \left[ \,2(u v-1) \, \right] ~~.
&(3.15) \cr} $$

	At first sight, our initial ans\"atze (3.2) and (3.3) appeared to be
trivial, because the metric in $~D=(2,2)$~ became {\it flat}.  However,
this is not trivial at all from the following
consideration closely related to the peculiar feature of the
SDSG system.  In the ordinary system with gravity in $~D=(1,3)$, we
can {\it not} always put the metric to be {\it exactly flat}, keeping other
matter fields propagating, due to the presence of the energy-momentum tensor of
the matter fields, which does {\it not} equate the vanishing Einstein tensor
in the left-hand side of the gravitational field equation.
This is physically equivalent to the fact that even a ``free'' field will
generate gravity around it.  If we review eq.~(2.6), we see that the vanishing
of the  first two terms for the flat metric does not pose this sort of problem,
because the remaining $~\Hat\phi\-$dependent terms can vanish, thanks to the
peculiar quadratic terms in $~\Hat\phi$~ or $~\Hat F$~
contracted by the $~\e^{A_1\cdots
A_8}\-$tensor, even when the field $~\Hat\phi$~ satisfies its free field
equation.\footnotew{This is also related to the fact that the self-dual
Riemann tensor is automatically Ricci-flat.}
In other words, in the SDSG system the scalar field will not create
any gravitational field around itself! This is the very feature that enables
the
SDSG system to  embed such $~D=2$~ dilaton gravity black hole solutions in a
non-trivial way,  and thus strongly suggests that the $~N=2$~ superstring
theory
is the  possible ``master theory" of such lower-dimensional systems.

We mention another possible class of solutions with non-vanishing
values for $~\Hat G\du{\hat\m\hat\n}{A B}$.  For example, instead of (3.7b) we
can put
$$\Hat G\du{m n}{A B} = \e_{m n} \, g^{A B}~~, ~~~~
\Hat G\du{a b}{A B} = \e_{a b} \, g^{A B}~~,
\eqno(3.16)$$
with  arbitrary constants $~g^{A B}$, such that the anti-self-duality (2.5) is
maintained.  Since $~\Hat G\du{\hat\m\hat\n}{A B}$~ appears in field equations
only in (2.10), this can provide a generalized class of non-trivial solutions
for the $~N=8$~ SDSG background.  The $~SO(8)$~ symmetry is {\it
global}, so that {\it any} of $~g^{A B}$~ can be non-zero constants,
with each component providing an $~U(1)$~ monopole-type solution.  We
stress that this special feature of non-zero $~\Hat
G\du{\hat\m\hat\n}{A B}$~ solutions without disturbing other field equations is
again peculiar to the $~N=8$~ SDSG.

\bigskip\bigskip\bigskip

\centerline{\bf 4.~~Exact Solutions of Dilaton with Gravitational Instanton}

	Our result so far suggests the possibility of other sets of
exact solutions including a dilaton field embedded into the $~{\bf
70}\-$dimensional scalars.  We now show that this is indeed
the case by presenting exact solution of the dilaton field on the
Eguchi-Hanson gravitational background [11].  To this end, we
temporarily forget about the dimensional reduction we have used so far,
and treat the system in the total $~D=(2,2)$.  Our result is very similar
to the exact solution given in ref.~[12], where the $~N=1$~ SDSG is coupled
to self-dual tensor multiplet and the self-dual Yang-Mills multiplet on
the Eguchi-Hanson background.

	A self-dual Lorentz connection for the Eguchi-Hanson metric [11] can
be easily obtained as [12]
$$ \eqalign{& \Hat\o\du 2 {(1)(2)} = \Hat\o\du 2{(3)(4)} =
\half \sqrtrf \sin\psi ~~, ~~~~
\Hat\o\du 2 {(1)(4)} = \Hat\o\du 2{(2)(3)} = - \half \sqrtrf \cos\psi ~~, \cr
& \Hat\o\du 3{(1)(2)}= \Hat\o\du 3{(3)(4)}=
- \half \sqrtrf \sinhth \cos\psi ~~, ~~~~
\Hat\o\du 3 {(2)(3)} =\Hat\o\du 3{(1)(4)}
= - \half\sqrtrf \sinhth\sin\psi ~~,  \cr
& \Hat\o\du 3 {(1)(3)} = \Hat\o\du 3{(2)(4)} = \half \left( 1 + \fracmm1{r^4}
\right) \coshth~~, \cr
& \Hat\o\du 4{(1)(3)}= \Hat\o\du 4 {(2)(4)}
= \half \left( 1 + \fracmm1{r^4} \right) ~~ , \cr }
\eqno(4.1) $$
This Lorentz connection can reproduce the self-dual Riemann tensor for the
Eguchi-Hanson metric [5,12]:
$$ \eqalign{ d s^2 = &\, \fracmm1{\onemi} d r^2 + \fracmm{r^2} 4
\left( 1 - \fracmm1{r^4} \coshthsq \right) d\varphi^2 - \fracmm14 r^2 d
\vartheta^2 + \fracmm{r^2}4 \left( \onemi \right) d\psi^2 \cr
& + \fracmm {r^2} 2 \left( \onemi \right)\coshth d\varphi d\psi ~~, \cr}
\eqno(4.2) $$
for our indefinite signature of $~D=(2,2)$ with the coordinate-labeling
$~(\Hat x^{\hat\m}) = (r, \vartheta, \varphi, \psi)$.

We next use the ansatz for the dilaton field similar to (3.6) embedded into
$~\phi\low{A B C D}$:
$$\Hat \phi\low{1234} \equiv \phi (r)~~,
\eqno(4.3) $$
only with the $~r\-$dependence.  For simplicity, we put all other fields
to be zero.  Eventually the only non-trivial field equation to be solved
is (2.9) for the dilaton
$$ \phi''(r) + \fracmm{3r^4+1}{r(r^4-1)} \phi'(r) = 0 ~~,
\eqno(4.4) $$
where each prime denoting the derivative $~d/d r$.
This equation is exactly the same as eq.~(3.17) of ref.~[12] for a dilaton
field coupled to the $~N=1$~ SDSG up to a non-homogeneous term from a $~\Hat
F\Hat F\-$term in the latter, and is solved by
$$ \phi = a \ln\left( \fracmm{r^2 - 1}{r^2+1} \right) + b ~~,
\eqno(4.5) $$
with arbitrary constants $~a,~b$.  The difference of this set of
exact solutions from the previous black hole solution is the absence of
the non-vanishing $~\Hat F_{\hat\m\hat\n A B}\-$field.

	Our result here suggests that the maximal $~N=8$~ SDSG has
other SDSG with fewer supersymmetries as its sub-theories.
We stress that this is just a simple example with the
dilaton field on non-trivial self-dual gravitational background, and
the universal feature of our methods used in this Letter can be applied to
get more interesting solutions embedded into the $~N=8$~ SDSG.

\bigskip\bigskip\bigskip

\centerline {\bf 5.~~Concluding Remarks}

In this paper we have shown that
the massless bosonic fields in $~N=8$~ SDSG, as the consistent background for
the $~N=2$~ closed superstring, has the $~D=2$~ dilaton gravity black
hole solution.  The interesting point is that the exact solution, which were
originally designed for the backgrounds in the bosonic or $~N=1$~ superstring,
turned out to be an important solution also for the $~N=2$~ superstring
theory.

We have utilized the simple dimensional reduction scheme with
appropriate truncation, to get a set of exact solutions for the
backgrounds of the $~N=2$~ superstring.  In our previous papers [13,14] we
have taken a separate approach to the dimensional reduction of $~N=2$~ {\it
open}  superstring to generate supersymmetric integrable models [13] and
topological field theories [14].  The results
in the present Letter have different interesting  aspects, due to the exact
solutions closely related to the bosonic and $~N=1$~ superstrings.  Our
result provides encouragement for the trial of finding other non-trivial
solutions as a generalization of the dilaton gravity black hole solution.

In our embedding, the dilaton field is embedded into one component of
the original $~{\bf 70}\-$dimensional representation of $~SO(8)$.  This is also
the reflection of the fact that in the $~N=8$~ SDSG there is {\it a priori} no
particular single scalar playing a role of dilaton, but all the scalars fit
into the irreducible $~{\bf 70}\-$representation.  Also to be stressed is the
interesting feature about the topological solution for the $~\Hat
F_{\hat\m\hat\n A B}$, which played an important role to supply the constant
term in the $~\phi\-$field equation (3.8).  We have also seen the
important aspect of the SDSG allowing the propagating $~\Hat\phi_{A B
C D}\-$field, while the metric is kept {\it exactly flat}.
It is interesting to note that
the black hole solution emerges out of a simple scalar field equation.

	We have also presented another set of exact solutions with a
scalar field embedded into $~\Hat\phi\low{A B C D}$~ on the Eguchi-Hanson
gravitational instanton background [11], which shares the same exact
solution as the dilaton field in what we call self-dual tensor
multiplet coupled to $~N=1$~ SDSG system [12].
This result suggests that the maximal $~N=8$~ SDSG contains other
SDSG as its sub-theories {\it via} appropriate truncations.

In regard to the dilaton field hidden among scalars in the maximal
SG, we can find a analogous situation in the usual $~D=(1,3),\, N=8$~
SG [15].  We have shown in our previous paper [15] that
two components out of the $~70$~ scalars in the usual $~N=8$~ SG in
$~D=(1,3)$~ can be separated from the other components systematically,
as the dilaton and an
axion fields.  Therefore, by following a procedure similar to ref.~[15], we may
be able to re-formulate the $~N=8$~ SDSG in a more suitable way, such that the
dilaton and axion fields are treated separately from the outset, at the expense
of the manifest global $~SO(8)$~ symmetry.

Even though we have dealt only with the bosonic field equations in this Letter,
we stress the important role played by supersymmetry in the $~N=8$~ SDSG.  The
peculiar $\Hat\phi\Hat F\Hat F\-$type coupling was required {\it only} by
supersymmetry, but not strongly motivated by any other purely bosonic self-dual
theory.  It is plausible that other coupling in the lagrangian (2.1) will
lead to more interesting exact solutions.

Our result in this Letter is natural from the viewpoint
that the $~N=2$~ superstring has great chance to be the ``master theory'' of
supersymmetric integrable models in lower-dimensions [2-6,13,14,16].
Our result gives other supporting evidence for this conjecture for the
{\it closed} $~N=2$~ superstring, in addition to the recent examples
[13,14,16] for {\it open} $~N=2$~ superstring.
This applies to any other possible exact solutions that may lead to
conformal field theories in $~D=2$~ [16], as well as topological field
theories in $~D=3$~ [14].  It is highly plausible that the $~N=2$~
superstring theory generate a wider class of other unknown solutions in
lower dimensions in such a comparatively easy way as by the free scalar
field equation we have seen.

The original mathematical conjecture [1] that all the
integrable systems are generated by $~D=(2,2)$~ self-dual Yang-Mills theory,
applied {\it only} to {\it bosonic} integrable systems.  The recent development
for  $~N=2$~ superstring [7,8] in physics has further promoted this conjecture
to a more general one that all the space-time {\it supersymmetric}
integrable systems in lower-dimension are generated by $~N=2$~
superstring.  It is amusing to note that the
structure of $~N=2$~ superstring is elaborate enough to generate
those supersymmetric integrable systems as the most fundamental underlying
``master theory''.

One of the motivation and advantage of embedding exact solutions into the
``master theory'' is the possibility of describing {\it
non-perturbative} deformation of integrable systems.
Suppose $~D=(2,2)$~ ``compactifies'' into $~\[ D=(2,0) \] \otimes \[E=(0,2)\]$,
where both manifolds are compact, such as the product of two tori:~$T^2 \otimes
T^2$, and consider the possible non-zero solution for $~\Hat
G\du{\hat\m\hat\n}{A B}$~ like (3.16).  Then the anti-self-duality condition
(2.5) for the $2\-$form $~\Hat G$~ yields  $$\int_{D=(2,0)}~ \Hat G = -
\int_{E=(0,2)} ~\Hat G~~,
\eqno(5.1) $$
where $~\Hat G$~ in both sides have the common $~{\scst A B}\-$components.
Interestingly (5.1) relates the two kinds of {\it quantized} $~U(1)$~ monopole
charges on $~D=(2,0)$~ and $~E=(0,2)$.  We can find many solutions
describing different vacua, which may be connected to each other by ``quantum
tunneling''.

In our recent paper [16], we have shown that the $~N=4$~ self-dual
supersymmetric Yang-Mills theory,  which is consistent background for {\it
open}
$~N=2$~ superstring, can  generate $~N=(1,1)$~ and $~N=(1,0)$~ WZNW models on
coset $~G/H$~ in  $~D=2$.  We have found that the supergeometry of a
superconformal WZNW  model has universal composite superpotential, which
satisfies the super-integrability condition $~F_{A B} = 0$~ in superspace
[13,14,16].  This result is also natural according to our past experience
that SG usually has the aspects of supersymmetric Yang-Mills theories
as its built-in features.  It is also plausible that the dilaton
gravity black hole solution can have a link to the
self-dual Yang-Mills theory {\it via} a $~\s\-$model based on the
coset $~SL(2,\IR)/U(1)$~ [9].  Considering these results, it seems that the
{\it open and closed} $~N=2$~ superstring theory is really playing key roles
for generating lower-dimensional integrable systems with or without curved
backgrounds.  We expect much more development in this direction related to the
self-dual supersymmetric systems in the future.

\bigskip\bigskip

We are indebted to S.J\.Gates, Jr.~and W\.Siegel for valuable suggestions.

\bigskip\bigskip
\vfill\eject

\refs
\small

\def\\{\vskip 0.05in}
\def\item#1{\Item{#1}}
\def\items#1{\\ \item{[{#1}]}}

\items{1} M.F.~Atiyah, unpublished;
\item{  } R.S\.Ward, Phil.~Trans.~Roy.~Lond.~{\bf A315} (1985) 451;
\item{  } N.J\.Hitchin, Proc.~Lond.~Math.~Soc.~{\bf 55} (1987) 59.

\items{2} S.V.~Ketov, S.J.~Gates, Jr.~and H.~Nishino, \pl{308}{93}{323}.

\items{3} H.~Nishino, S.J.~Gates, Jr. and S.V.~Ketov,
\pl{307}{93}{331}.

\items{4} S.J.~Gates, Jr., H.~Nishino and S.V.~Ketov,
\pl{297}{92}{99}.

\items{5} S.J.~Ketov, H.~Nishino and S.J.~Gates, Jr., \np{393}{93}{149}.

\items{6} H.~Nishino, Maryland preprint, UMDEPP 93--79, to appear
in Int.~Jour.~Mod.~Phys.

\items{7} H.~Ooguri and C.~Vafa, \mpl{5}{90}{1389};
\np{361}{91}{469}; \ibid{367}{91}{83};
\item{  } H.~Nishino and S.J.~Gates, Jr., \mpl{7}{92}{2543}.

\items{8} W.~Siegel, \pr{47}{93}{2504}.

\items{9} E.~Witten, \pr{44}{91}{314};
\item{  } G.~Mandal, A.M.~Sengupta and S.R.~Wadia,
\mpl{6}{91}{1685};
\item{  } T.~Eguchi, \mpl{7}{92}{85}.

\items{10} J.~Scherk and J.H.~Schwarz, \np{153}{79}{61}.

\items{11} T.~Eguchi and A.~Hanson, \ap{120}{79}{82};
\item{   } E.~Calabi, Ann.~Sci.~Ec.~Norm.~Sup.~{\bf 12} (1979) 269.

\items{12} H.~Nishino, \pl{307}{93}{339}.

\items{13} S.J.~Gates, Jr.~and H.~Nishino, \pl{299}{93}{255};
\item{   } H.~Nishino, \pl{318}{93}{107};
\item{   } H.~Nishino, Maryland preprint, UMDEPP 94-47 (Oct.~1993).

\items{14} H.~Nishino, \pl{309}{93}{68}.

\items{15} S.J.~Gates, Jr.~and H.~Nishino, \cqg{8}{91}{809}.

\items{16} H.~Nishino, \pl{316}{93}{298}.

\end{document}